\documentclass[letterpaper,twocolumn,10pt]{article}
\usepackage{usenix,epsfig,endnotes}
\usepackage{booktabs}
\usepackage{graphicx}
\usepackage{subfigure}
\usepackage{enumerate}

\begin{document}


\title{\Large \bf pCAMP: Performance Comparison of Machine Learning Packages\\ on the Edges }

\author{
{\rm Xingzhou Zhang, Yifan Wang}\\
Wayne State University\\
 Institute of Computing Technology, CAS\\
University of Chinese Academy of Sciences
\and
{\rm Weisong Shi}\\
Wayne State University
} 



\maketitle

\pagestyle{empty}

\subsection*{Abstract}
Machine learning has changed the computing paradigm. Products today are built with machine intelligence as a central attribute, and consumers are beginning to expect near-human interaction with the appliances they use.
However, much of the deep learning revolution has been limited to the cloud.
Recently, several machine learning packages based on edge devices have been announced which aim to offload the computing to the edges. However, little research has been done to evaluate these packages on the edges, making it difficult for end users to select an appropriate pair of software and hardware.
In this paper, we make a performance comparison of several state-of-the-art machine learning packages on the edges, including TensorFlow, Caffe2, MXNet, PyTorch, and TensorFlow Lite. We focus on evaluating the latency, memory footprint, and energy of these tools with two popular types of neural networks on different edge devices. 
This evaluation not only provides a reference to select appropriate combinations of hardware and software packages for end users but also points out possible future directions to optimize packages for developers.
\footnote{This paper has been published by USENIX HotEdge 18. 
	
	Contact Email: xingzhou92@gmail.com}

\section{Introduction}\label{sec:introduction}
Recently, the burgeoning field of Machine Learning based applications has been enabled by the advances in three directions: ML models, processing power, and big data. 
In the cloud, developing new machine learning models needs not only numerous computational resources but also much time. In order to improve the efficiency of training models, many open-source machine learning packages have been recently developed, including TensorFlow \cite{abadi2016tensorflow} from Google, Caffe \cite{jia2014caffe} from UC Berkeley, CNTK \cite{seide2016cntk} from Microsoft, PyTorch \cite{collobert2011torch7} from Facebook, and MXNet \cite{chen2015mxnet} supported by AWS.

In the next five years, increasingly more data will be generated at the edge of the network collected by the Internet of Things, which in turn calls for edge computing \cite{Shi2016EdgeVision}. 
Keeping data processing on the edges (such as edge servers, home server, vehicles, and mobile phones) is effective in offering real-time services \cite{satyanarayanan2015edge,zhang2018icdcs,lee2017gremlin,qi2017vehicle}. However, different from the data center in the cloud, edge nodes (also known as edge servers or fog nodes), usually have very limited computing power. Edge nodes are not a good fit for training ML-based models since they require a large footprint in both the storage (as big as 500MB for VGG-16 Model \cite{simonyan2014very}) and computing power (as high as 15300 Million Multiply-Adds \cite{howard2017mobilenets}).

To support executing large-scale models on the edges, several machine learning packages based on edge devices have been announced (Caffe2 for mobile \cite{Caffe2}, TensorFlow Lite \cite{TensorFlow-Lite}) which aim to dramatically reduce the barrier to entry for mobile machine learning. Unlike the training process on the cloud, machine learning models on the edges have been trained on the cloud, and the packages are designed to execute inference on the edges.


\begin{table*}[!htbp]
\centering  
\caption{The overview of the combinations of hardware and software packages.} 
\label{table:overview}
\begin{tabular}{lccccccc}
\toprule
 & MacBook Pro& Intel FogNode &NVIDIA Jetson TX2 & Raspberry Pi&Nexus 6P \\
\midrule
TensorFlow&      $\surd$&$\surd$ &$\surd$&$\surd$&$\times$\\
Caffe2&          $\surd$&$\surd$ &$\surd$&$\times$&$\times $\\
MXNet&           $\surd$&$\surd$ &$\times$&$\times$&$\times$\\
PyTorch&         $\surd$&$\surd$ &$\surd$&$\times$&$\times$\\
TensorFlow Lite& $\times$&$\times$&$\times$&$\times$&$\surd$\\
\bottomrule
\end{tabular}
\end{table*}

However, few studies have evaluated these packages on edge devices.
In this paper, we make a performance comparison of several state-of-the-art machine learning packages on the edges, including TensorFlow, Caffe2, MXNet, PyTorch, and TensorFlow Lite. We focus on their performance when running a trained model on the edges rather than the training process.
We evaluate the latency, memory footprint, and energy of these packages with a large-scale model, AlexNet \cite{krizhevsky2012imagenet}, and a small-scale model, SqueezeNet \cite{iandola2016squeezenet}, on the edges. The devices are MacBook Pro (MacBook) \cite{MacBook}, Intel's Fog Reference Design (FogNode) \cite{fognode}, NVIDIA Jetson TX2 (Jetson TX2) \cite{jetsontx2}, Raspberry Pi 3 Model B+ (Raspberry Pi) \cite{raspberrypi}, and Huawei Nexus 6P (Nexus 6P) \cite{nexus6p}.


Our major insights are summarized as follows: (1) There are no obvious single winners on every metric of all hardware platforms, but each has its merits. (2) The time required to load models is greater than it takes to run the model for most packages, which implies that there exist some opportunities to further optimize the performance. (3) Sacrificing space to improve execution efficiency can be a good method to optimize these packages.

The remainder of the paper is organized as follows. Section 2 presents the related work, and Section 3 introduces our experimental methods. Results and discussions are presented in Section 4. We make a conclusion and discuss our future work in Section 5.

\begin{table*}[!htbp]
\centering  
\caption{The experimental setup of hardware.} 
\label{table:hardware setup}
\begin{tabular}{@{}llllllll@{}}
\toprule
                  & CPU    & Frequency & Cores & Memory &       OS                     \\ \midrule
MacBook Pro       & Intel Core i5         & 2.7GHz    & 4     & 8GB    & macOS 10.13.2                    \\
FogNode    & Inter Xeon E3-1275 v5 & 3.6GHz    & 4     & 32GB   & Linux 4.13.0-32-generic          \\
Jetson TX2 & ARMv8 + NVIDIA Pascal GPU & 2GHz & 6     & 8GB    & Linux 4.4.38-tegra    \\
Raspberry Pi      & ARMv71                & 0.9GHz    & 4     & 1GB    & Linux rasberrypi 4.4.34               \\
Nexus 6P   & Qualcomm Snapdragon 810 & 2GHz   &  8     & 2.6GB  & Android 8.0 3.10.73              \\
\bottomrule
\end{tabular}
\end{table*}

\section{Related work}\label{sec:Background and Related work}

In 2016, Shi \cite{shi2016benchmarking} evaluated the performance of Caffe, CNTK, TensorFlow, and Torch and saw how they performed on cloud platforms, including two CPU platforms and three GPU platforms. This work focused on evaluating the training speed of these tools with fully connected neural networks (FCNs), convolutional neural networks (CNNs), recurrent neural networks (RNNs), and benchmarking some distributed versions on multiple GPUs. 
Robert Bosch LLC presented a comparative study of five deep learning frameworks, namely Caffe, Neon, TensorFlow, Theano, and Torch, on three aspects: extensibility, hardware utilization, and speed \cite{bahrampour2015comparative}. 
In 2017, Louisiana State University analyzed the performance of three different frameworks, Caffe, TensorFlow, and Apache SINGA, over several HPC architectures in terms of speed and scaling \cite{shams2017evaluation}. It focused more on the performance characteristics of the state-of-the-art hardware technology for deep learning, including NVIDIA's NVLink and Intel's Knights Landing.

All of the research summarized above focused more on evaluating performance on the cloud. To our knowledge, our work is the first concerning the performance of machine learning packages on the edges.

\section{Experimental Methods}\label{sec:Experimental Methods}

\subsection{Machine learning packages}
Based on the popularity of these packages, this paper evaluates the performance of TensorFlow, Caffe2, MXNet, PyTorch, and TensorFlow Lite (as is shown in Table \ref{table:overview}). Keras is also widely used; since it is built on top of TensorFlow, so we do not consider it. 

TensorFlow \cite{abadi2016tensorflow} is developed by Google which has integrated most of the common units into the machine learning framework. It is an open source software library for numerical computation using data flow graphs.  TensorFlow Lite is a lightweight implementation of TensorFlow for mobile and embedded devices.
It enables on-device inference with low latency and a small binary size.

Caffe \cite{jia2014caffe} is a deep learning framework made with expression, speed, and modularity in mind. It is developed by Berkeley AI Research (BAIR) and by community contributors. Caffe2 is a deep learning framework that provides an easy and straightforward way to experiment with deep learning and leverage community contributions of new models and algorithms. 

MXNet \cite{chen2015mxnet} is a flexible and efficient library for deep learning. It was initially developed by the University of Washington and Carnegie Mellon University, to support CNN and long short-term memory networks (LSTM). 

PyTorch \cite{collobert2011torch7} is published by Facebook. It is a python package that provides two high-level features: tensor computation with strong GPU acceleration and deep Neural Networks built on a tape-based auto-grad system.

\subsection{Machine learning models}

Recently, convolutional neural networks (CNNs) have been deployed successfully in a variety of applications, including ImageNet classiﬁcation \cite{krizhevsky2012imagenet}, face recognition \cite{lawrence1997face}, and object detection \cite{zbontar2016stereo}. 
Since the pre-trained AlexNet and SqueezeNet are widely available on many frameworks, we will use these two CNN-based models, AlexNet as the large-scale model and SqueezeNet as the small-scale model, to do evaluations.

AlexNet competed in the ImageNet Large Scale Visual Recognition Challenge in 2012 \cite{LSVRC}. 
It has 61,100,840 parameters and 240MB size. 
Currently it is often used for evaluating the performance of hardware and software.


SqueezeNet \cite{iandola2016squeezenet} was proposed in 2016 as a small DNN architecture. It achieves AlexNet-level accuracy on ImageNet with 50x fewer parameters. It has 1,248,424 parameters and 4.8MB size.


\subsection{Edge devices}

As is shown in Table \ref{table:hardware setup}, this paper adopts different kinds of hardware, including MacBook, FogNode, Jetson TX2, Raspberry Pi, and Nexus 6P. Their CPUs, memory, and operating systems are also different. 
Note that Jetson TX2 is an embedded AI computing device, it contains an ARM CPU and a NVIDIA Pascal GPU.

Table \ref{table:overview} lists the combinations of hardware and software packages that we were able to install successfully and get the experimental results in this paper. It is a big challenge to install these packages on different types of hardware. 
In order to ensure the fairness of the evaluation, the trained AlexNet and SqueezeNet models that we used were downloaded from the official websites (such as Caffe2 Model Zoo \cite{Caffe2ModelZoo}). For TensorFlow, we only adopted AlexNet to do the experiments because the official website has not yet published a trained SqueezeNet model. For TensorFlow Lite, neither SqueezeNet nor AlexNet were officially available, so for TensorFlow Lite on Nexus 6p, we ran a pre-trained MobileNet \cite{howard2017mobilenets} model.



\subsection{Metrics}

This paper evaluates the performance of packages when executing the inference tasks on the resource-constrained edges. Therefore, it leverages latency, memory footprint, and energy as the metrics.


This paper measures latency by using the ``time'' function in Python source scripts. Latency is divided into two metrics: inference time and total time. 
When packages are running the trained models, they should import the package first, and then load the model and image. Then it calls the function provided by the package to do the inference. To avoid errors, we called the inference function 100 times in each experiment. The inference time is the average time of the 100 inferences take. The time that the whole process takes is the total time.

For Nexux 6P, we monitor the memory footprint by leveraging the ``Trepn'' application developed by Qualcomm \cite{Trepn-Profiler}.
For Linux-based systems, which include FogNode, Jetson TX2, and Raspberry Pi, we get the memory footprint from the ``ps'' instruction. 
For MacBook, the ``Activity Monitor'' application provides each process's memory footprint.

\begin{figure}[h]
	\centering
	\includegraphics[width=2.5in]{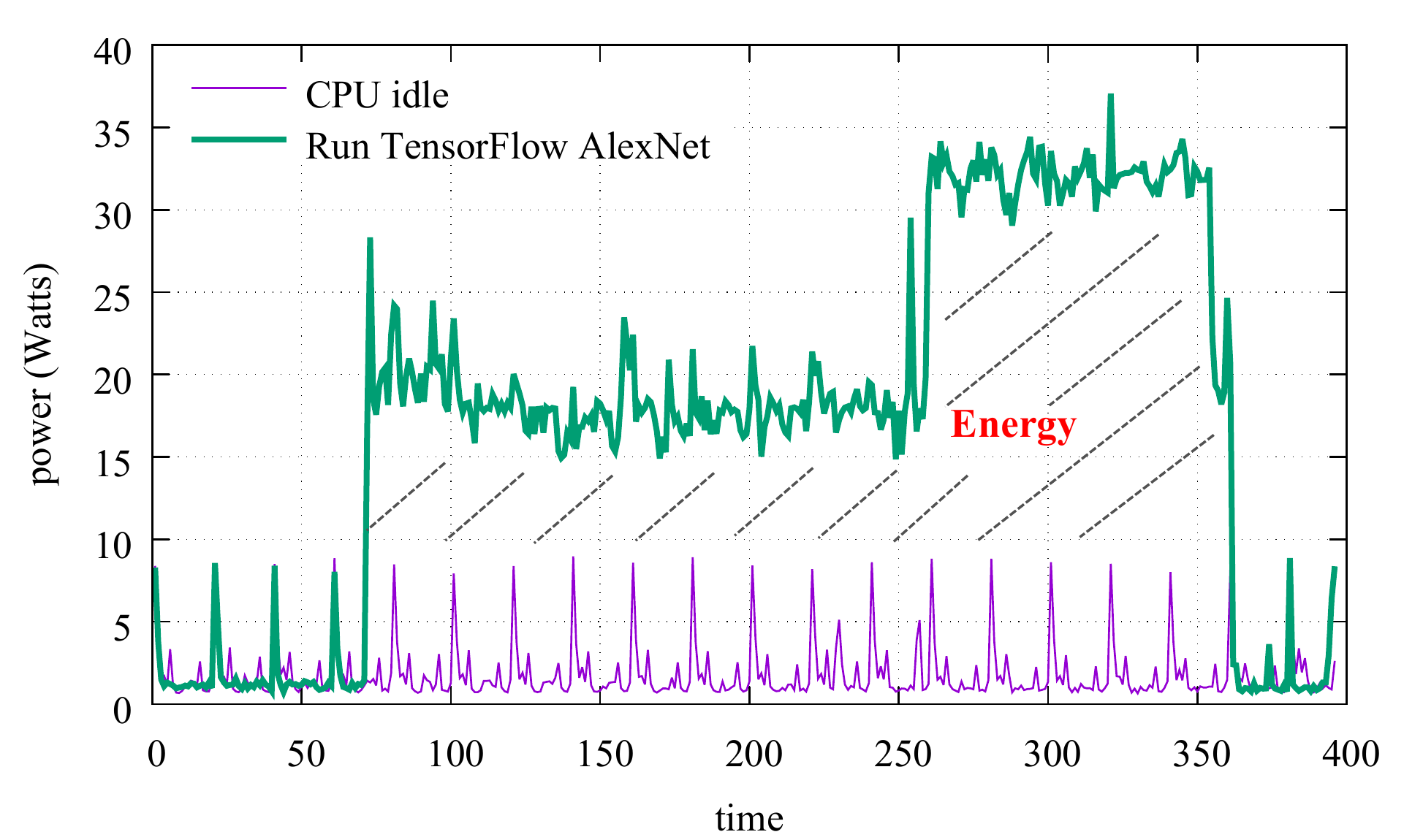}
	\caption{The method to obtain energy.}
	\label{fig:energy_metrics}
\end{figure}

For Nexus 6p, the energy can be obtained by leveraging the ``Trepn'' application.
For the Linux-based operating system, we get the energy from the RAPL\cite{desrochers2016validation} tools on Linux. The power usage of each process cost can be monitored by RAPL.
For MacBook, we use the ``Intel Power Gadget'' application to get the energy \cite{IntelPowerGadget}. The app estimates the power/energy usage of all processes. To get the energy that just the package costs, we get a baseline first. We stop almost all service and wait until the power usage of the CPU is stable. Then we collect the power usage for 400 seconds. As is shown in Figure \ref{fig:energy_metrics}, the purple line represents the power usage when the CPU is idle, which is the baseline. Then, we start a process (such as run a pre-trained model) and collect the power usage also for 400 seconds. The green line shown in Figure \ref{fig:energy_metrics} represents the power consumed by the process. 
Then we can estimate the energy that the process costs by calculating the area crossed by the two lines, as shown by the shaded part in Figure \ref{fig:energy_metrics}.

\section{Results and Discussions}\label{sec:Results}

\begin{figure*}[!htb]
    \centering
    \subfigure[\label{fig:Inference time}Inference time]{\includegraphics[height=1.8in]{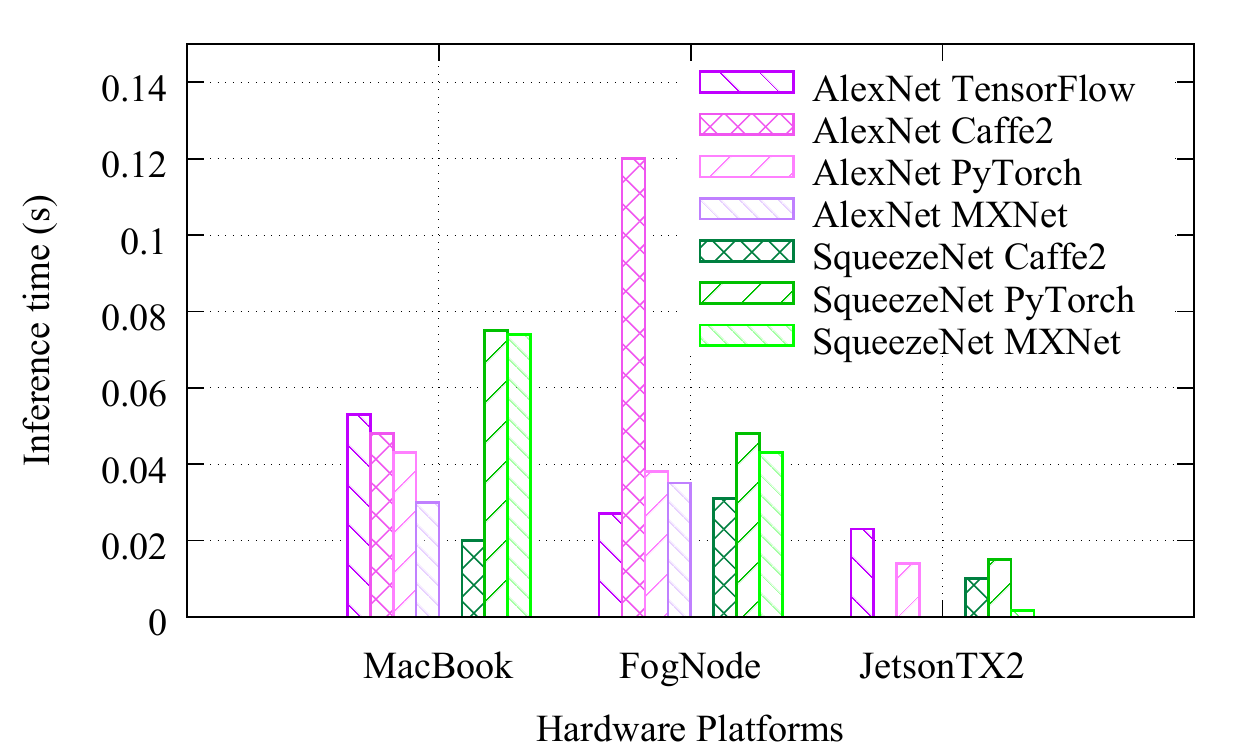}}
    \subfigure[\label{fig:Total time}Total time]{\includegraphics[height=1.8in]{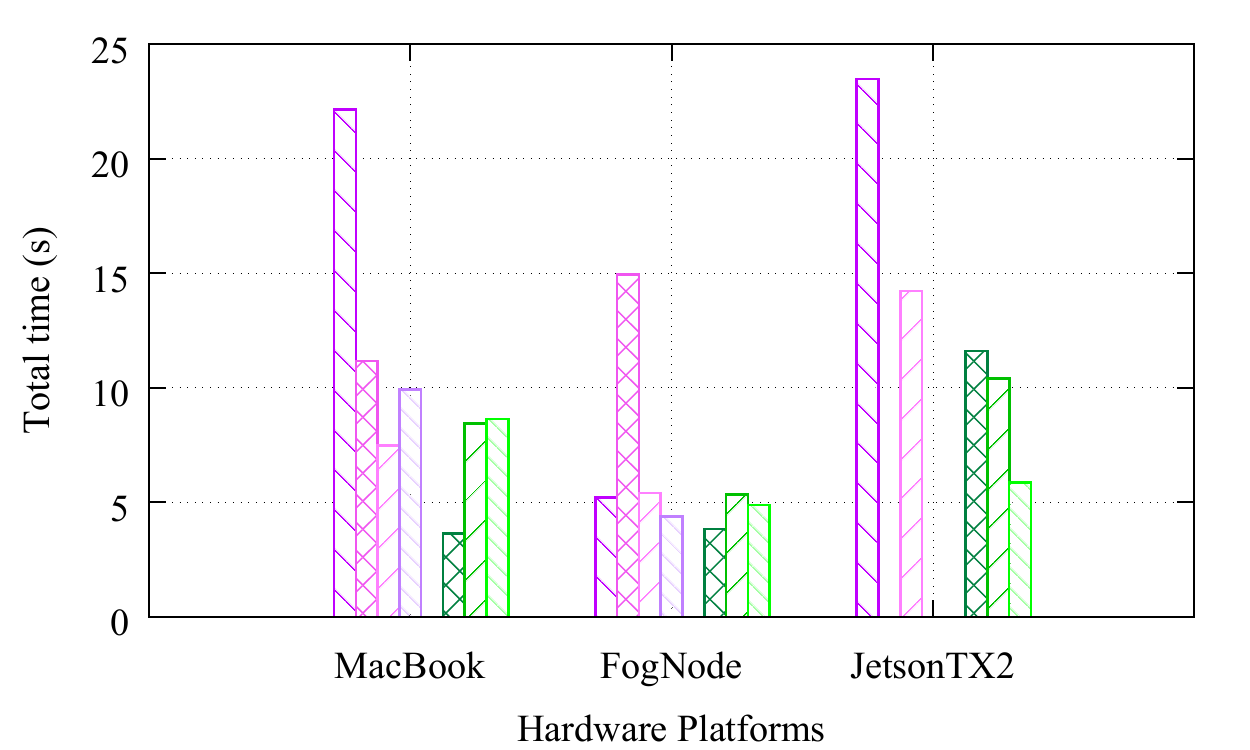}}
    \subfigure[\label{fig:Memory footprint}Memory footprint]{\includegraphics[height=1.8in]{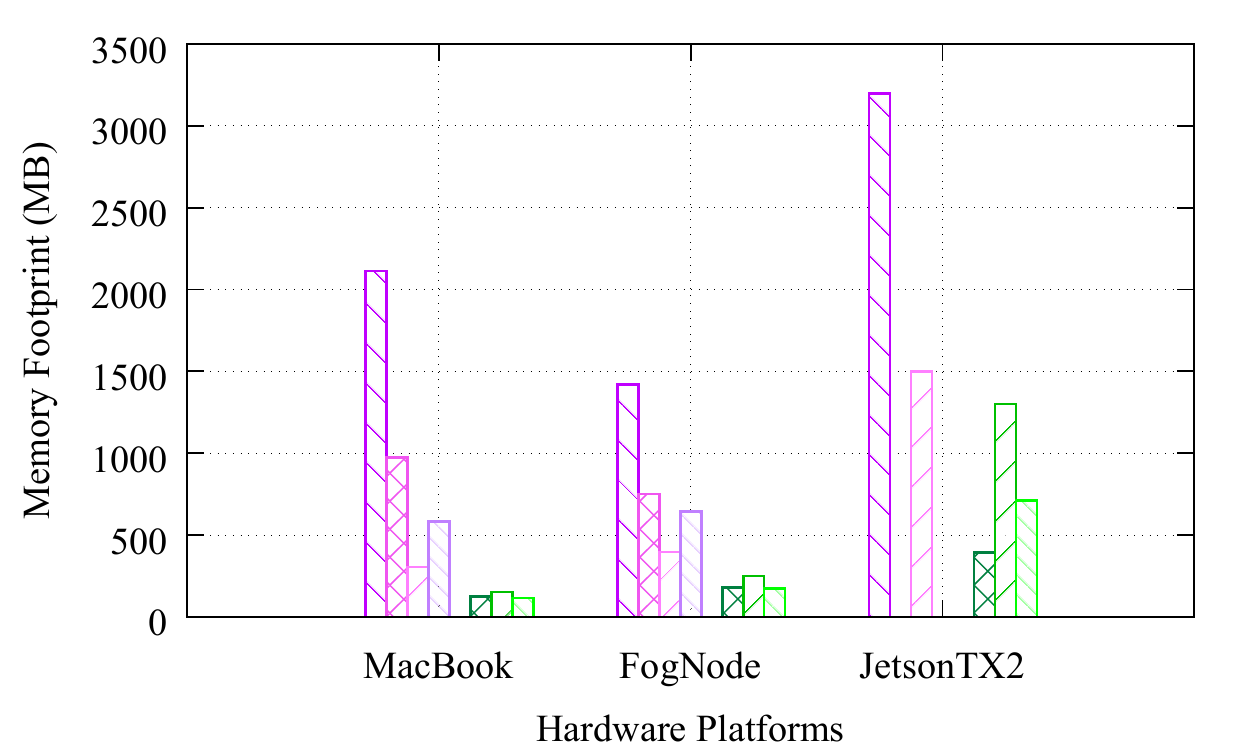}}
    \subfigure[\label{fig:Energy}Energy]{\includegraphics[height=1.8in]{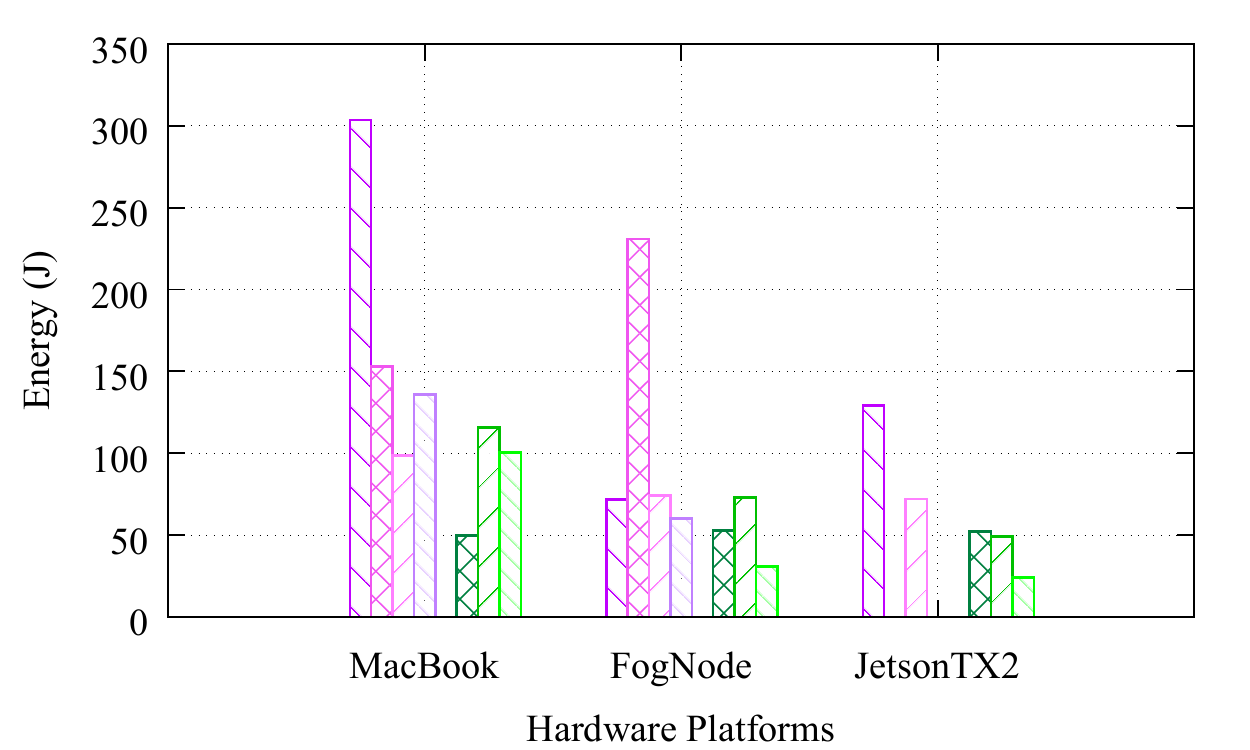}}
    \caption{Performance comparison of Machine Learning Packages on the Edges}
    \label{fig:performance comparison}
\end{figure*}

Figure \ref{fig:performance comparison} shows the results of MacBook, FogNode, and Jetson TX2. The results on Raspberry Pi and Nexus 6P will be presented directly in the text. 
The four figures have the same legends. Because of space limitations, we just plot the legend in Figure \ref{fig:Inference time} and omit others. 
As is explained in Section \ref{sec:Experimental Methods}, since the official website of TensorFlow does not provide trained SqueezeNet, we do not benchmark TensorFlow on SqueezeNet.

\subsection{Latency}
On AlexNet, as is shown in Figure \ref{fig:Inference time} and Figure \ref{fig:Total time}, MXNet has in general the best performance on MacBook. It only takes 0.029s to execute an inference task, which is 1/2 of TensorFlow's 0.053s. On FogNode, both PyTorch (0.038s) and MXNet (0.035s) perform slightly worse than TensorFlow (0.027s). 
Caffe2's performance on FogNode and Jetson TX2 are not good. The inference time of Caffe2 is around 4x longer than TensorFlow on FogNode. Due to install limitation, we only compare the performance of TensorFlow and PyTorch on Jetson TX2 when running AlexNet, PyTorch has a shorter inference time and total time than TensorFlow.
On SqueezeNet, Caffe2 results in the best performance whose inference time is only about 1/4 of PyTorch and MXNet on MacBook. MXNet works slightly better than PyTorch. On FogNode, they have the similar latency. Each package takes less time on Jetson TX2 than MacBook and FogNode, especially for MXNet. The packages use optimized GPU code and so run faster on the Jetson TX2, the only system with a supported GPU.


Due to the resource constraints on Raspberry Pi, the inference time when running SqueezeNet on Caffe2 is 2.08s, and the total time is 218.67s. The inference time of Nexus 6P when running MobileNet based on TensorFlow Lite is 0.26s. Although it is much longer than running on MacBook, FogNode, and Jetson TX2, it is really great progress for running a model on small edge devices.

\begin{table}[h]
\centering
\caption{Time profile of PyTorch on Jetson TX2}
\label{table:time profiling}
\begin{tabular}{lcc}
\toprule
                    & AlexNet (s) & SqueezeNet (s) \\ \midrule
import package              & 0.77            & 0.74                            \\
load model          & \textbf{9.50}            & \textbf{6.20}                          \\
first load image    & 0.12              & 0.15                           \\
first inference     & \textbf{2.34}            & \textbf{1.76}                           \\
second load image   & 0.075              & 0.07                             \\
second inference    & 0.006              &0.019                             \\
inference 100 times & \textbf{1.35}            & \textbf{1.45}                           \\
total time            & 14.17            & 10.41                           \\ \bottomrule
\end{tabular}
\end{table}

To profile where the time is spent, we divide the whole process of inference into seven parts and collect the time that each part takes.
Table \ref{table:time profiling} represents the time of every part takes when it leverages PyTorch to execute the inference task on Jetson TX2. Doing the inference 100 times only needs 1.35s while loading model needs 9.50s when running AlexNet. It has similar performance when running SqueezeNet. In addition, we also find that doing the inference for the first time takes longer than the other 100 times for both AlexNet and SqueezeNet.

\textbf{Insight 1:}
The time taken to load models is greater than what it takes to run the model for some packages, which implies that there exist some opportunities to further optimize the performance on the edges by decreasing the loading latency and frequency.

\subsection{Memory footprint}

On AlexNet, as is shown in Figure \ref{fig:Memory footprint}, the memory footprints on MacBook and FogNode are similar. TensorFlow occupies more than 2000MB memory, which is the largest. PyTorch performs the best as it has the least memory utilization, less than 500MB. On JetsonTX2, the memory footprint of TensorFlow is up to 3000MB while PyTorch utilizes only 1/2 of TensorFlow.

On SqueezeNet, the three packages have the similar memory use on CPU-base hardware, MacBook and FogNode. On Jetson TX2, PyTorch used 1301MB memory, which is 2x of MXNet (709MB) and 4x of Caffe2 (375MB). 
The memory of Caffe2 is 132MB when running SqueezeNet on Raspberry Pi. TensorFlow Lite occupies about 84M memory when running MobileNet on Nexus 6P. 

Analyzing Figure \ref{fig:performance comparison}, we find that these packages favor tradeoff between memory and latency. On AlexNet, the inference time of TensorFlow is much shorter than Caffe2 on FogNode, while it utilized more memory. On SqueezeNet, the total time of MXNet is 2x than Caffe2 while MXNet occupies 1/2 memory of Caffe2.

\textbf{Insight 2:}
Based on the observation, we believe that MXNet may sacrifice space to improve efficiency, which can be a useful optimized method for executing real-time tasks on the edges. 

\subsection{Energy}

\begin{figure}[h]
    \centering
    \includegraphics[height=1.5in]{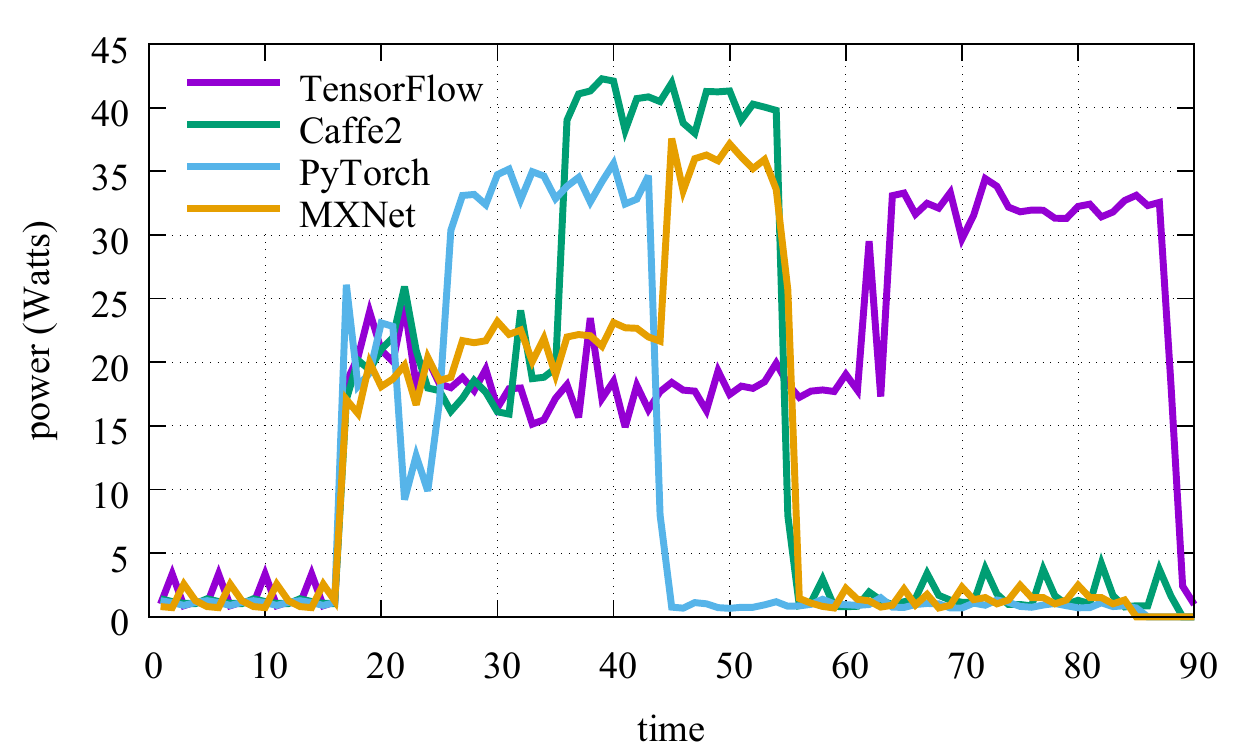}
    \caption{Energy profiling of AlexNet on FogNode}
    \label{fig:Energy profiling}
\end{figure}

On MacBook, Caffe2 could be regarded as the most energy-efficient package as it consumes the minimal energy. On FogNode, Caffe2 consumes about 2x energy of other packages when running AlexNet. PyTorch performs better than TensorFlow on JetsonTX2. MXNet has the best performance on FogNode and JetsonTX2 when running SqueezeNet.
We take running AlexNet on FogNode as an example to profile the energy. Figure \ref{fig:Energy profiling} shows the energy consumption of these packages. Based on the energy consumption, each process can be roughly divided into two periods: load models and inference. The power of the loading model is smaller than the inference, but the time that the loading model takes is greater than the inference for TensorFlow and MXNet.

\textbf{Insight 3:}
Comparing Figure \ref{fig:Memory footprint} and Figure \ref{fig:Total time}, we can draw a conclusion that there is a positive correlation between energy and latency. Longer latency leads to a greater energy consumption.

\subsection{Summary}
Figure \ref{fig:performance comparison} shows that there are no obvious single winners on every metric among the hardware platforms; each has its merits. 
We summarize the merits of every package and provide a reference to select appropriate packages for end users:

\begin{enumerate}[1)]
\item TensorFlow is faster when running large-scale model on the CPU-based platform FogNode, while Caffe2 is faster when running small-scale model. MXNet is the fastest package when running on Jetson TX2.
\item PyTorch is more memory efficient than other packages for MacBook and FogNode.
\item MXNet is the most energy efficient package on FogNode and Jetson TX2 while Caffe2 performs better on MacBook.
\end{enumerate}





\section{Conclusion}\label{sec:Conclusion}
This paper aims to evaluate the performance of a set of machine learning packages when running trained models on different edge hardware platforms. By comparing the latency, memory footprint, and energy of these packages on two neural network models, AlexNet and SqueezeNet, this paper finds that there are no single packages that can consistently outperform others in both time and space dimensions. Also, the process of loading model takes more time for some packages than running the model, which implies that there exist some opportunities to further optimize the performance on the edges.

In the future, we plan to benchmark some optimized inference-only frameworks (such as Intel Computer Vision SDK \cite{IntelComputerVisionSDK}) and more hardware platforms (such as iPhone and Movidius Myriad VPU chip \cite{VPU}).
Also, we will try to propose a unified metric that can combine different metrics to evaluate these packages on the edges.


\newpage
{\footnotesize \bibliographystyle{unsrt}
\bibliography{ref}}

\end{document}